\documentclass[aps,pre,groupedaddress,showpacs]{revtex4}
\usepackage{amsmath} 
\usepackage{graphicx}
\usepackage{amssymb}
\usepackage{psfrag,float} 
\usepackage{subfigure}

\begin{document} \bibliographystyle{apsrev} 
\title{Synchronization of 
oscillators with long range interaction: phase transition and anomalous finite 
size effects}
\author{M\'at\'e Mar\'odi}
\email{marodi@complex.elte.hu}
\author{Francesco d'Ovidio}
\email{dovidio@fysik.dtu.dk}
\author{Tam\'as Vicsek}
\email{vicsek@angel.elte.hu}
\affiliation{$^*\ddag$Dept. of Biological Physics, E\"otv\"os University, 
Budapest, P\'azm\'any P. Stny. 1A, 1117 Hungary\\
$\dag$Chaos Group and Quantum
Protein Center, Dept. of Physics, 
Building 309, Danish Technical University, DK-2800 Lyngby, Denmark} 
\date{\today}

\begin{abstract}

Synchronization in a lattice of a finite population of phase oscillators 
with algebraically decaying, non-normalized coupling is
studied by numerical 
simulations. A critical level of decay is found, below which 
full locking takes place if the population contains 
a 
sufficiently large number of elements. For large number of 
oscillators and small coupling constant, numerical simulations and analytical 
arguments 
indicate that a phase transition separating synchronization from incoherence 
appears at a decay exponent value 
equal to the number of dimensions of the lattice. 
In contrast with earlier results on similar systems with normalized coupling, 
we have indication that for the decay exponent less than the dimensions of the 
lattice and for large populations, synchronization is
possible even if the coupling is arbitarily weak. This finding suggests
that in organisms interacting through slowly
decaying signals like light or
sound, collective oscillations can always be established if the population is
sufficiently large.

\end{abstract}

\pacs{05.45.Xt}
\maketitle

\section{Introduction} \label{sec:intro} 
Synchronization is one of the most 
fascinating nonlinear phenomena appearing in a wide range of fields. Examples 
cover physical systems (network of Josephson junctions \cite{josephson1}), 
oscillating chemical reactions \cite{kuramotobook}, molecular turnover in
allosteric enzymes \cite{stange98}, a variety of biological 
observations (synchronous flashing of fireflies \cite{buck}, menstrual 
synchrony \cite{mcclintock}, metabolic activity in yeast cells \cite{dano99,
dano02}, 
and other various physiological processes \cite{glass}), and social phenomena 
(rhythmic applause \cite{clap}).

Models for rhythmic synchronization use populations of nonlinearly coupled 
oscillators \cite{winfree67,winfree80,pikovsky01}. These models might be 
grouped according to the nature of coupling. For instance, for pulsatile 
interaction integrate-and-fire oscillators provide a useful tool of 
description \cite{mirollostrogatz} while in the case of continuous interaction 
a system of coupled limit cycle oscillators has proved to be a good model 
\cite{kuramotobook}.

An important aspect of synchronizing systems is the spatial dependence of
the 
coupling between two oscillators. Two limit cases have been deeply 
investigated: i) mean field (with each pair of oscillators interacting with a 
given coupling strength, independent of their position, ii) oscillators on a 
lattice with interactions only between nearest neighbours. However, realistic 
systems are quite different. In many of the biologically relevant systems the 
signals carrying information about the phase of an oscillator decay as a 
function of the distance slowly. For example, the intensity of the sound or 
light signals decays as a power law, in particular, in the three dimensional 
case as the inverse of square of the distance from the source. 
In Ref. \cite{rogers96} results have been presented concerning the
existence of a critical exponent below which 
full synchronization can be achieved. These results considered a coupling term 
normalized by the sum of the spatial coefficients and the limit of the 
population size going 
to infinity.

However, in a real system typically the interaction 
between two oscillators depends only on the spatial distance and the phase 
differences and thus there is no normalization coefficient. This change, 
trivial at first sight, introduces a peculiar dependence of the dynamics on 
the number of oscillators, and thus 
places the study of \emph{finite} populations into relevant context.

In this paper synchronization in populations of phase 
oscillators with a non-normalized coupling term is studied. Our
investigations are especially focused on 
the 
dependence of full locking and the mean field on the number of oscillators for 
different value of the exponent $\alpha$ describing the spatial decay of the 
interactions. The main result of our work is the identification (through 
numerical simulations and a simple analytical argument) of critical values of 
$\alpha$ below which full 
synchronization, or in more general, collective oscillations can take
place if 
the population is sufficiently large. We remark that very few works in our 
knowledge has approached the problem of finite size populations.
 
The paper is organized as follows. In Sec. \ref{sec:kur} we summarize
previous results on the versions of the model of our studies, the
Kuramoto model.
In Sec. \ref{sec:size} we explore 
the boundary of full locking in the plane of the decay exponent $\alpha$ 
and the number $N$ of oscillators, at different 
coupling values. We find that if $\alpha$ is below a critical value 
$\alpha_c(K)$ full synchronization can be achieved if the system is 
sufficiently large, i.e., above a critical size $N$. We argue however that 
this information is partial for real systems, since bulk oscillations can take 
place also if the system is not completely locked. We thus study in Sec.\ 
\ref{sec:trans} the behaviour of the mean field. It appears then that a 
phenomenon 
close 
to a thermodynamic phase transition takes place looking at the mean field 
changing $\alpha$, with a discontinuity 
developing for $N$ going to infinity and increased fluctuations around the 
transition point. For $K$ going to zero, analytical arguments and numerical 
simulations indicate that the 
transition point occurs at $\alpha$ equal to the number of dimensions of the 
lattice. To show the robustness and generality of the 
result, we finally study a system of phase oscillators where the diversity is 
given by thermal noise. In such system full locking is not strictly defined, 
but on the contrary a mean field approach can be carried out in the 
same way as for oscillators with fixed parameter distribution and the same 
kind of phase transition appears. The system with thermal noise is very 
similar to the Heisenberg model with ferromagnetic interaction. We shall
suggest that a thermodynamic formalism 
may be applied to study analytically the phenomenon discussed numerically in 
this paper.

We remark that the present model has been proposed for real populations
of  oscillators (among others, specific examples are the above mentioned works
on Josephson junctions \cite{josephson1} and synchronized applause
\cite{clap}). In these cases, and when spatial dependences are included, our 
results have direct implications. They 
suggest that whenever the coupling signals of these systems decay slower than 
the  number of spatial dimensions (like for example light or sound), 
\emph{collective oscillations always arise if the population is sufficiently 
large, however weak the coupling is}.

\section{The Kuramoto model: previous results} \label{sec:kur} 
The original form of the Kuramoto 
model was introduced to describe oscillating chemical reactions 
\cite{kuramoto75,kuramoto1}. Later it proved to be useful in modeling a wide 
variety of processes (see e.g. Ref. \cite{clap2}). The basic concept is a 
population of $N$ interacting oscillators coupled via nonlinear 
phase-difference minimizing interaction. In general the equation of motion
of 
the phase $\phi_i$ of the $i$-th oscillator is: 
\begin{equation} 
\dot\phi_i=\omega_i+\sum_j \Gamma_{ij}(\phi_j -\phi_i), \label{eq:kur1} 
\end{equation} 
where $\omega_i$ is the natural frequency of the $i$-th 
oscillator, $\Gamma(\phi)$ is the two-oscillator interaction and the sum goes 
over a suitably defined subset of the population depending on the actual 
model. The system can be further simplified choosing, without loss of 
generality, a frame of reference rotating at the average natural frequency 
$\overline\omega_i=\omega_0$ and rescaling the time and the
coupling 
constant by the 
inverse of the standard deviation of the natural frequencies. The distribution 
of natural frequencies becomes then centered in zero and its standard 
deviation normalized.

Macroscopic states of this system can be characterized by a real \emph{order 
parameter}: \begin{equation} z(t)=\left | \frac 1N \sum_{j=1}^N 
e^{\textrm{i}\phi_j (t)} \right |. \label{eq:op} \end{equation}
Another approach to synchronization is the investigation of frequency-locked 
states, i.e. when all $\dot \phi_i(t)$ approach asymptotically the same value.

Two basic setups of the Kuramoto model have been studied extensively so far. 
First, let us consider the case when

\begin{equation} \Gamma_{ij}(\phi)=\frac KN \sin(\phi), \end{equation} which 
defines a \emph{mean-field interaction}, where $K>0$ is the coupling
strength.  Now the summation in \eqref{eq:kur1} goes over the whole
population. Let us  assume that the distribution of the natural frequencies
$\omega_i$ is  symmetric about 0, and is convex in that point. Under these
conditions the  existence of a critical coupling $K_c$ was proved, above which
partial synchronization is possible  \cite{kuramoto2,strogatz00, demonte02}.
With the above conditions the frequency of the locked subset is equal to the
mean of the natural frequency distribution.

The other well explored case is when the interaction is inherently local, i.e. 
\emph{nearest neighbours} are coupled. Then the model takes the form: 
\begin{equation} \dot\phi_i=\omega_i+K\sum_{j\in \langle i \rangle} 
\sin(\phi_j-\phi_i), \label{eq:kurnn} \end{equation} where the symbol 
$\sum_{j\in \langle i \rangle}$ denotes summing over all the nearest 
neighbours of the $i$-th oscillator. Strogatz and Mirollo proved that 
the probability of a phase-locked solution 
tends to zero as the number of oscillators in the system goes to infinity 
\cite{strogatz1}. Also, they studied the more general case when the mean 
interaction exerted on one oscillator is uniformly bounded and the 
distribution of natural frequencies ``sufficiently broad-banded'' (for details 
see Ref. \cite{strogatz1}). For these conditions they showed that in the $N\to 
\infty$ limit the probability of a phase-locked state is zero.

An interesting situation arises when the interaction decays in space as a 
power function:

\begin{equation} \dot\phi_i=\omega_i+\frac K\eta\sum_{j\ne 
i}\frac{1}{r_{ij}^{\alpha}}\sin(\phi_j-\phi_i), \label{eq:al1} \end{equation}
where $r_{ij}$ is the distance between the $i$-th and $j$-th oscillator and 
$\eta$ is a normalization coefficient. It is clear that Eq.\ (\ref{eq:al1}) is
equivalent to the mean-field case when  $\alpha=0$. Also, when
$\alpha\to\infty$, the model becomes the same as in the  nearest neighbours
case. So, changing the exponent $\alpha$, one can have a  continuous
transition between the two extremes. Knowing the result of full 
synchronization in the global coupling case for high $K$, and only local 
synchronization in the nearest neighbours coupling for any $K$, one may
expect  a critical value of $\alpha$ below which full locking may be achieved.
This  was shown by Rogers and Wille, who studied Eq.\ (\ref{eq:al1}) setting 
$\eta=2\sum_{j=1}^{(N-1)/2} 1/j^\alpha$ in the limit of $N$ going to infinity
\cite{rogers96}. They demonstrated that critical values of $\alpha$ exist
depending on the coupling  strength $K$. A similar result holds for
sine-circle maps as well  \cite{plateau}. It was also recently shown
numerically that an adequate normalization $\eta$ removes the dependency of
the exponent $\alpha$ and system size $N$ on the fraction of oscillators
synchronized in frequency in the case $\alpha<d$ \cite{bahiana02}.

However, in most real systems the coupling between two oscillators
does not depend on global information on the population, but only  on the
phases and the distance of the two oscillators. It is thus realistic  for
these cases to set the normalization coefficient to 1. As an important 
consequence, the behaviour of the oscillators then strongly depends on  the
size of the population.

Throughout this paper we study models where
$\eta=1$. Before describing any numerical results, we remark that an 
estimation of a 
critical point can be obtained looking at Eq.\ (\ref{eq:al1}) and considering 
the coupling term. It is then natural to expect that the point $\alpha=d$, 
where $d$ is the number of dimensions of the lattice, has 
to play a special role. In fact if $\alpha\geq d$ the coupling term is 
bounded, for every $N$:

\begin{equation}\label{eq:bound} \left|\sum^N _{j\neq 
i}\frac{1}{r_{ij}^{\alpha}}\sin(\phi_j-\phi_i) \right|\leq \sum_{j\neq 
i}^{\infty}\frac{1}{r_{ij}^{\alpha}}<\infty. \end{equation}
However, for $\alpha\leq d$, in the limit of large $N$ the value of the 
coupling term
may entirely dominate the difference in natural frequencies $\omega_i$.
More precisely, there are regions of the phase space where this term may
diverge for $N$ going to infinity, in particular, in those regions where the difference
between phases is large.
Hence, we expect that for $\alpha\leq d$ synchronization is enhanced enlarging 
the size of the system, independently of the (positive) value of the
coupling 
constant $K$. This property and the
effects concerning the population size result from setting the coupling
without normalization, and follow directly from the divergence of 
the upper bound in relation
Eq.\ (\ref{eq:bound}).

\section{Finite population size and full synchronization boundaries} 
\label{sec:size}

In this Sec. we explore the effect of changing the number of oscillators in a 
one-dimensional system, measuring the boundary of the full locking region in 
the ($\alpha$, $N$) plane, for different $K$ values. From the brief discussion 
in the previous Sec. we may expect that for sufficiently low values of $K$ 
synchronization can be achieved if $N$ is sufficiently large, only if 
$\alpha<1$. On the contrary, this should not be possible for $\alpha>1$.

In order to identify the exact value of $\alpha$ at which the transition 
occurs it is helpful to consider the fraction $p$ of the oscillators that are 
locked at the main frequency $\omega_0$. We thus count how many oscillators 
asymptotically satisfy the condition:

\begin{equation} \dot\phi_i=\omega_0, \end{equation}
and we divide this number by $N$. 

The plane ($N$, $\alpha$) has been scanned in either 
vertical or horizontal direction, numerically integrating population of 
oscillators at those parameter values. The transition points were identified 
as the boundary of the region of full synchronization, i.e., measuring 
the transient $p$ and finding the parameter values where $p$ is no longer
equal to 
1. Refinements up to $\Delta 
\alpha=0.01$ and $\Delta N=1$ were used. 
In all numerical computations in this paper the natural frequencies $\omega_i$ 
were chosen from a Gaussian distribution with expectation value
$\overline\omega_i=0$ and 
unit variance.
Since the value of $N_c$ can in 
principle depend on the position of the natural frequencies on the lattice, 
each point has been averaged on several configurations, especially for low 
values of $N$. We remark that, setting without loss of generality the main 
frequency $\omega_0$ to zero, the locked state appears as a steady state. To 
locate it, a direct integration method (Euler) has been 
used. This method is simple to implement and reliable, since Eqs.\ 
(\ref{eq:al1}) are clearly non-stiff, but requires a careful setting of the 
parameters for computing $p$: in fact, the line $N_c$ corresponds to 
bifurcation points, where the transient time becomes more than exponentially 
long. Other methods, like continuation, could give a better estimation of the 
line but may be difficult to apply for large systems, having to deal with 
matrices of the order of $N^{2d}$.

\begin{figure}[H] \centering
\includegraphics[width=.7\textwidth,angle=0]{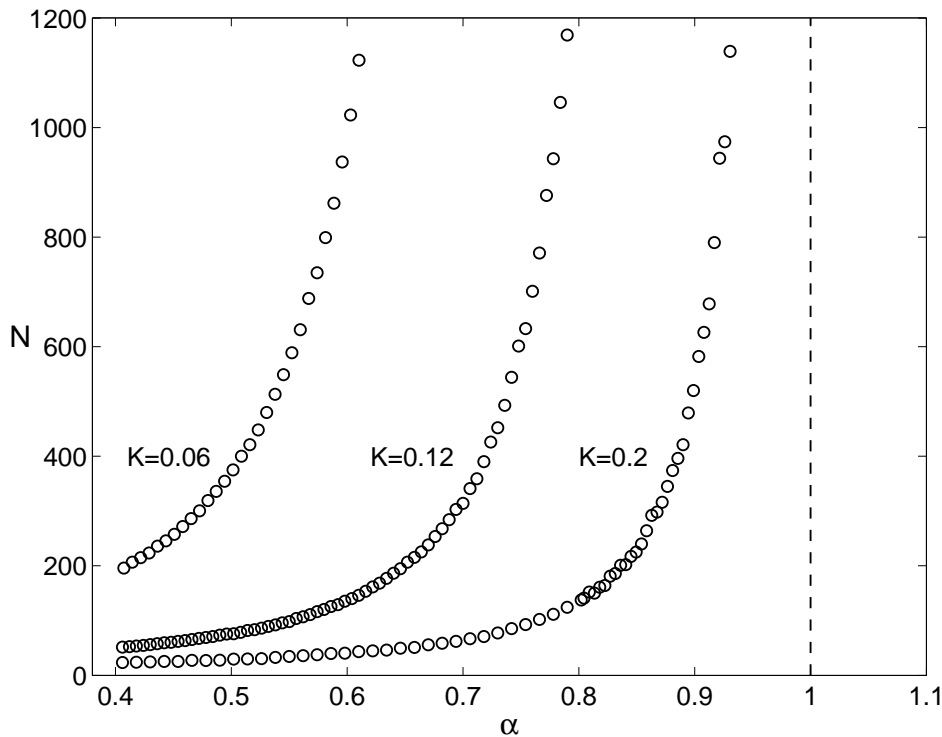} 
\caption{Boundary of the full locking region in the ($N$, $\alpha$) plane for 
different values of $K$. The line $\alpha=1$ is also shown.} \label{fig:bound} 
\end{figure}

For high coupling, full synchronization can appear at every $\alpha$ values. 
However, if $K$ is low we expect synchronization only for $\alpha$ values 
below the dimensions of the lattice. Results are plotted in Fig.\ 
\ref{fig:bound}.  The region 
$\alpha>1$ is outside the boundary of full synchronization, confirming the 
discussion of Sec.\ \ref{sec:intro}. The situation is more complex for 
$\alpha<1$. Here the effect of reducing the coupling can be indeed
compensated 
by increasing the number of oscillators, up to a critical value of $\alpha$. 
Such critical value however seems to be in general different from 1 (the 
number of dimensions of the lattice), and 
depending on the coupling strength $K$. This is due to the fact that for small 
$K$ values, synchronization is still enhanced by a larger number of 
oscillators, although the system is not completely locked. The next Sec. will 
investigate this situation, looking at the behaviour of the mean field.

\section{Collective oscillations} \label{sec:trans}

The results of the previous Sec. may be relevant for real systems. They 
suggest that in the case of oscillators coupled with a slowly decaying signal, 
full synchronization can be achieved as soon as the population is sufficiently 
large. However, the information that they give, describes only partially what 
may appear macroscopically. In this Sec. we propose to approach the problem 
looking at the behaviour of the mean field when $\alpha$ is changed. This 
criterion is less sharp than full synchronization: a transition cannot be seen 
for finite $N$ but, in analogy with thermodynamic phase transitions, looking 
at 
discontinuities for $N$ going to infinity. Nevertheless, as we shall see, the 
order parameter approach provides a robust and meaningful framework for 
describing the relations between synchronization and the decay of the 
coupling. Moreover, it allows to state a result in a simple way, connecting 
the critical value of the decaying of interaction with the number of 
dimensions of the lattice.

Let us consider numerical 
simulations conducted on one and two dimensional
lattices. We integrated 
the system of coupled differential equations \eqref{eq:al1} using the 
Euler method, timesteps varied from 0.01 to 0.0005 relative time-units
(measured in 
comparison with the frequencies).  The value of the order parameter is computed
calculating at 
each time $z(t)=|1/N\sum_{j=1}^Ne^{i\phi_j(t)}|$, 
discarding the first few thousand steps 
of integrations as transient.
Averaging this value on a long time (typically, 20000 time steps) gives
the measure of synchronization we use, $Z=\langle z(t) \rangle$.

\begin{figure}[H] \centering
\includegraphics[width=.7\textwidth,angle=0]{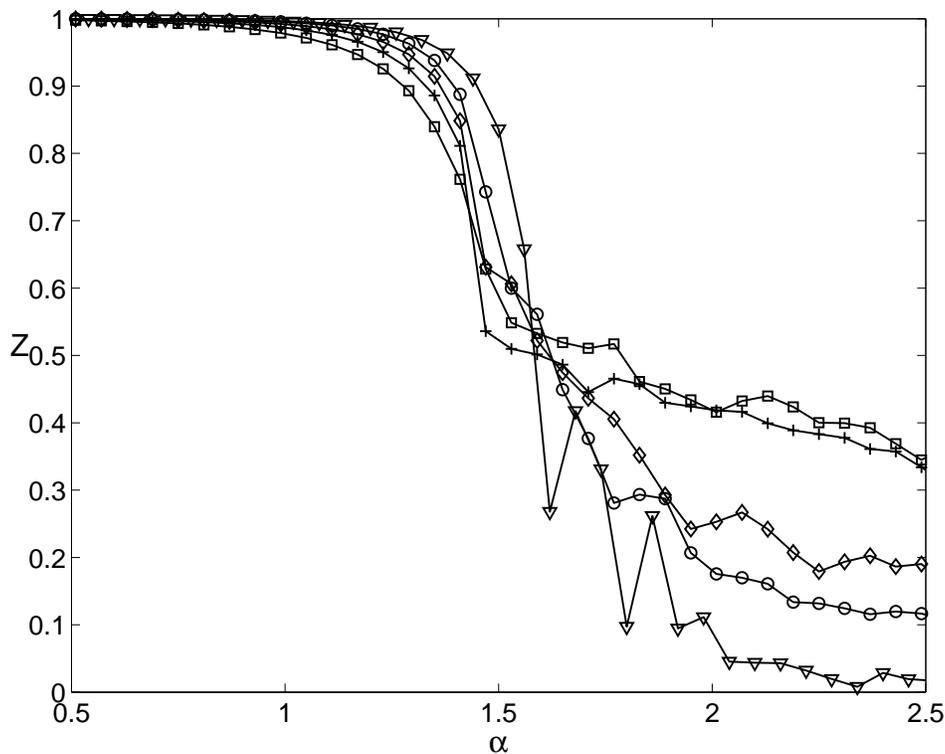} 
\caption{Time average of the order parameter as a function of $\alpha$ in 
$d=1$. Simulation results for $N=50, 100, 200, 500, 1000$ (symbols
are squares, pluses, diamonds, circles, and down triangles
respectively). The coupling is $K=1.0$.} \label{fig:az1d} \end{figure}

\begin{figure}[H] \centering 
\subfigure[]{\includegraphics[width=.4\textwidth,angle=0]{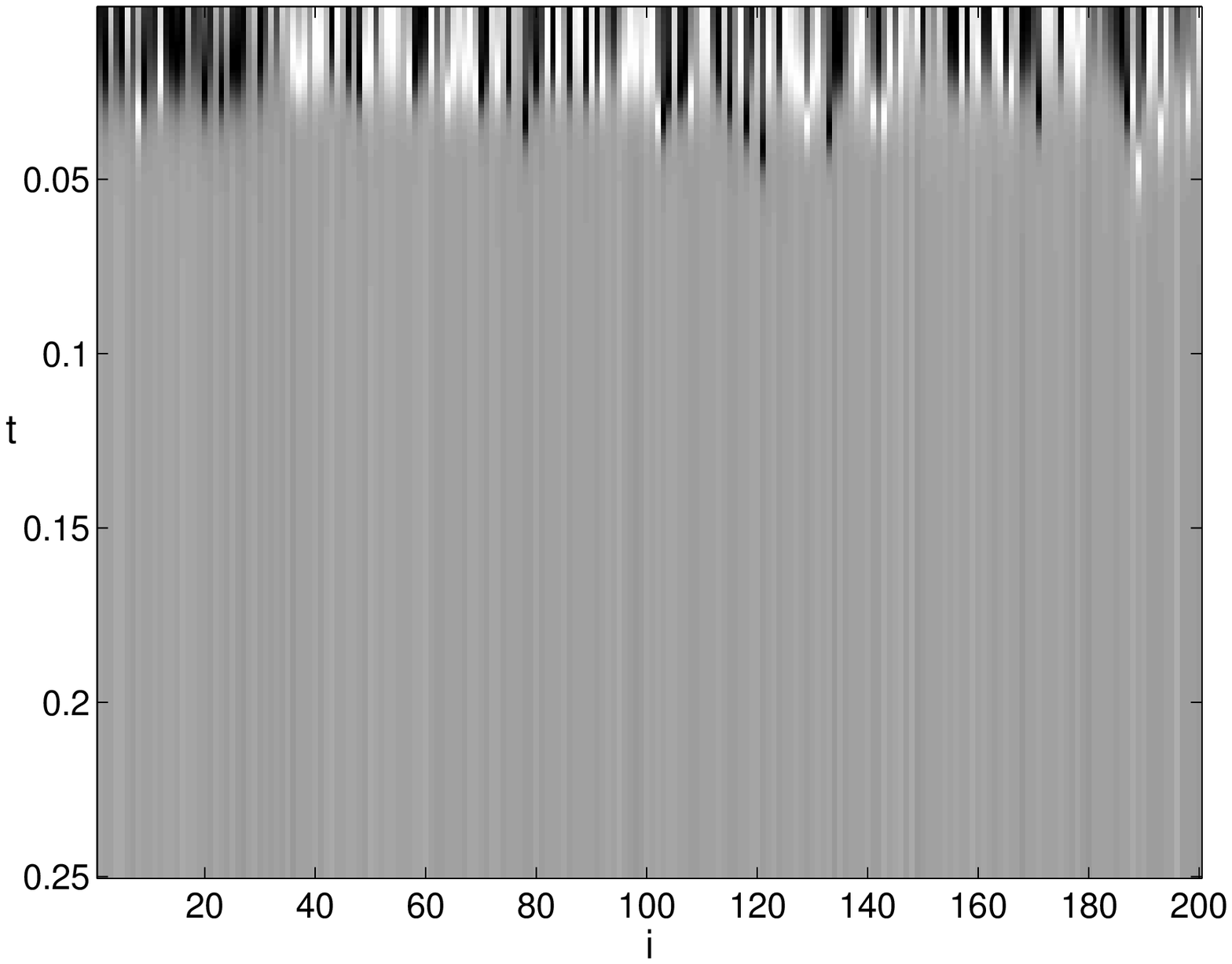}} 
\subfigure[]{\includegraphics[width=.4\textwidth,angle=0]{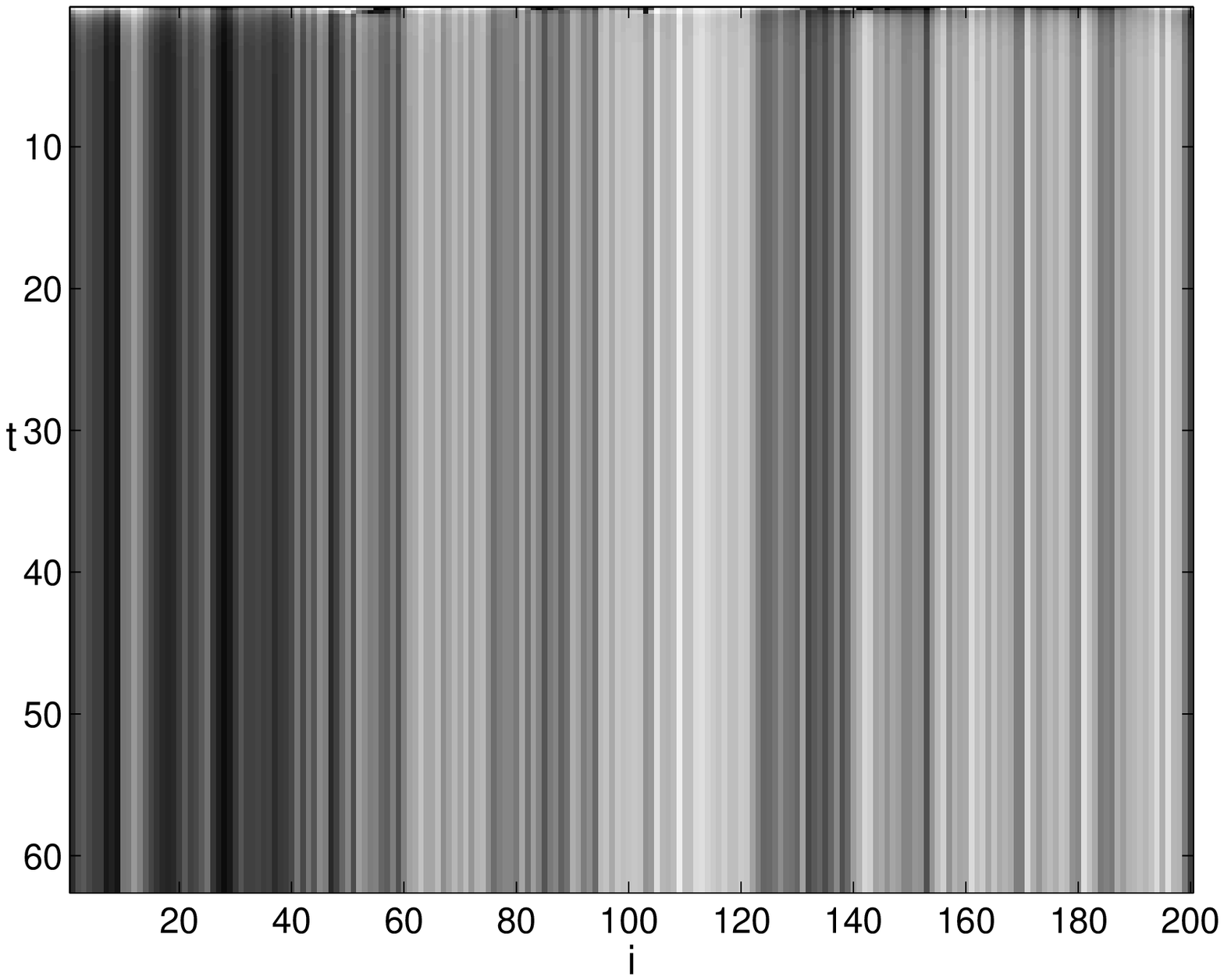}} 
\subfigure[]{\includegraphics[width=.4\textwidth,angle=0]{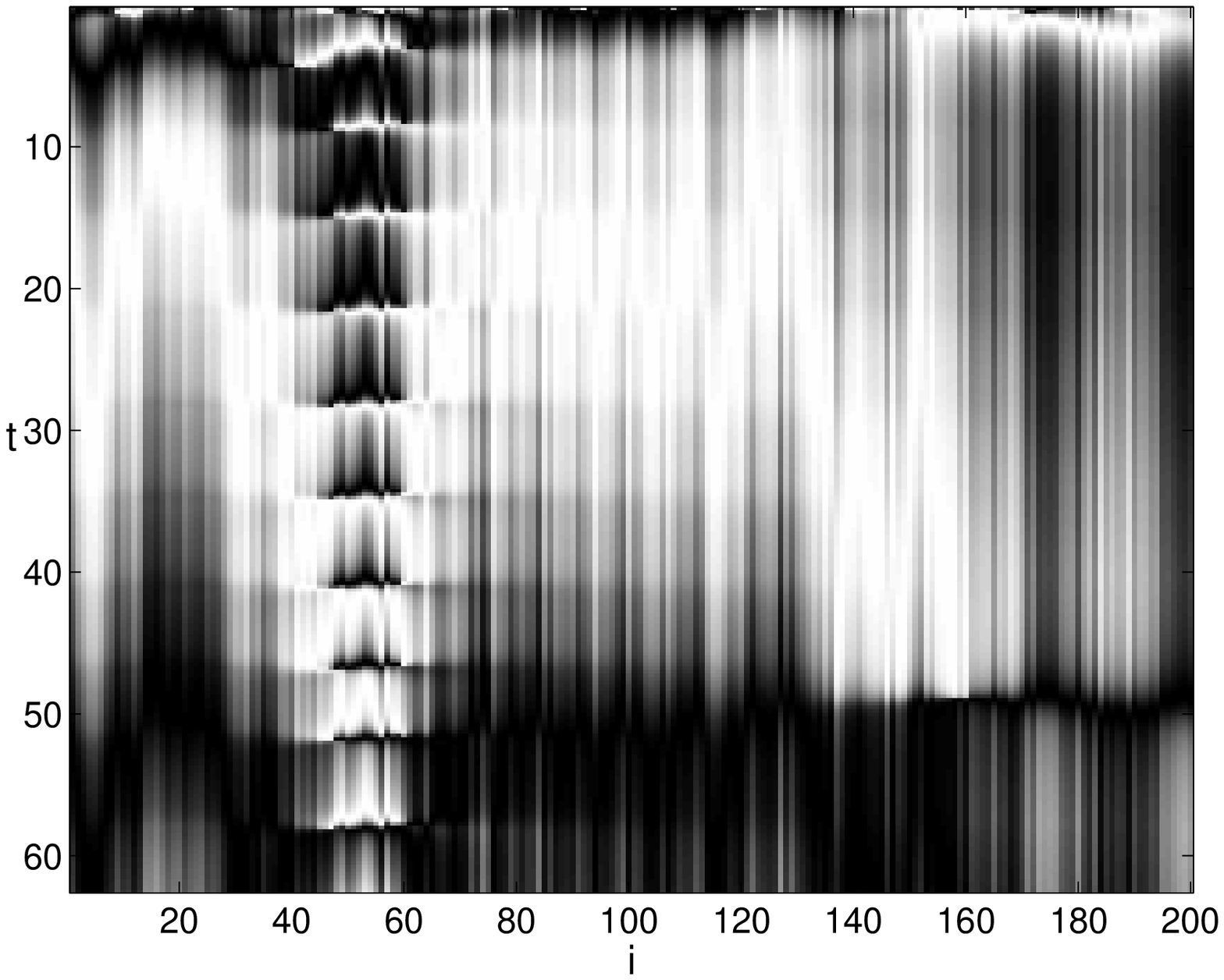}}
\subfigure[]{\includegraphics[width=.4\textwidth,angle=0]{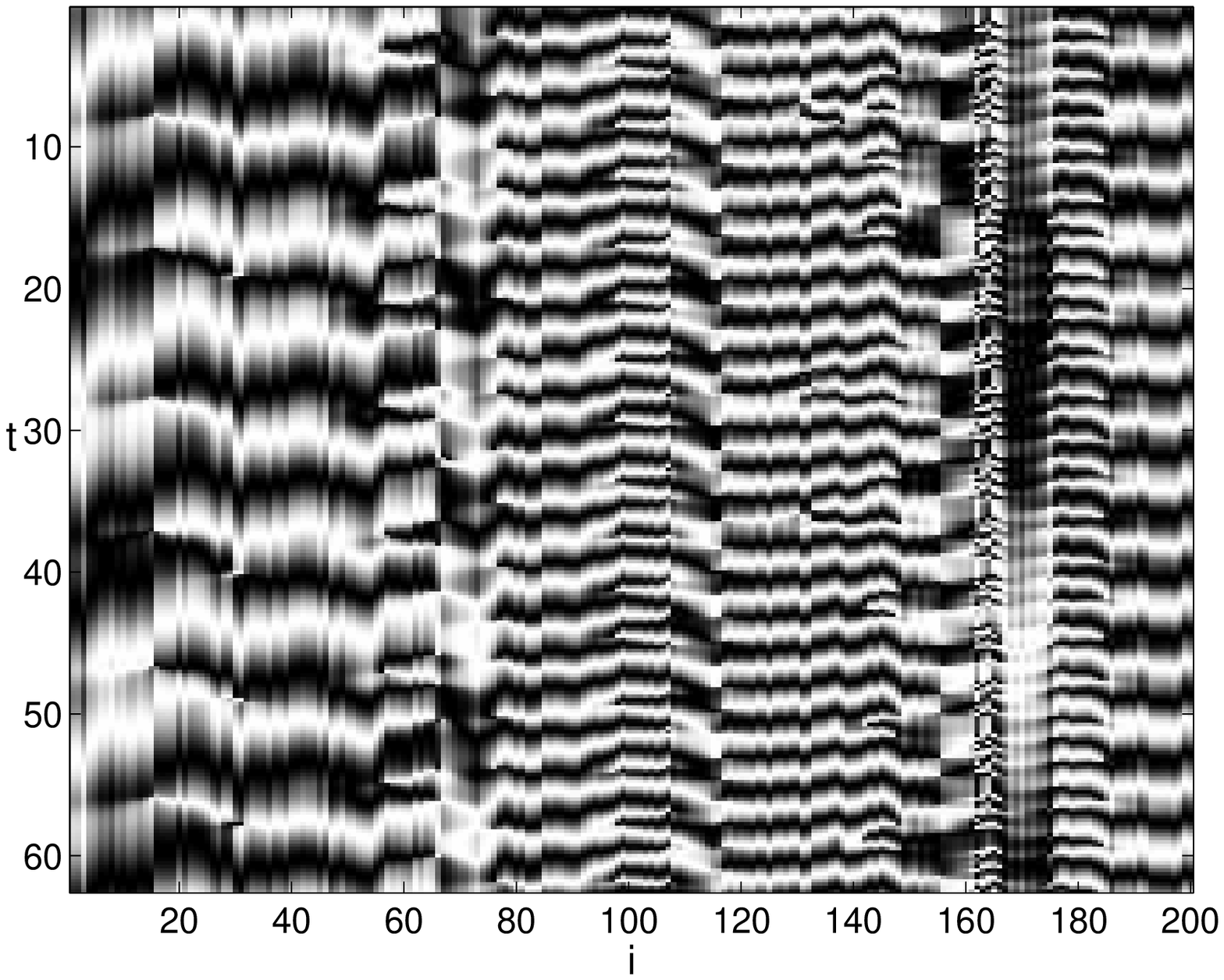}}
\caption{Space-time 
plots representing the sine of the phase of the $i$-th oscillator for different $\alpha$ 
values: (a) $\alpha=0.5$, (b) $\alpha=1.5$, (c) $\alpha=1.8$ and (d) 
$\alpha=2.5$. Greyscale indicates phases $\sin(\phi_i)$ from $-1$
(white) to $1$ (black). Other parameters were set to 
$K=1.0$ and $N=200$.} 
\label{fig:st} \end{figure}

One-dimensional results are plotted on Fig. \ref{fig:az1d}, where one can 
observe the transition from highly synchronized states at low $\alpha$ values 
to unsynchronized states. The behaviour of the system can be divided into 
three regimes. For $\alpha \lessapprox 1$ the system is fully locked, while 
for $\alpha \gtrapprox 2.5$ the order parameter $Z$ approaches a steady
value. 
In between one finds the region of transition. These three regions represent 
three different microscopic behaviors corresponding to the different 
macroscopic states (see Fig. \ref{fig:st}).
When $\alpha$ is low (e.g., $\alpha=0.5$) the system rapidly reaches the state 
of complete synchronization, when for all oscillators their frequencies
become 
$\dot\phi_i =\omega_0=0$. 
The next two
subfigures represent the transitional region ($\alpha=1.5, 1.8$). Here one can 
observe clusters arising first with increasing phase differences, then with 
synchronization holding only for finite times. This phenomenon leads to large 
fluctuations in the order parameter. Finally, in the unsynchronized region 
($\alpha=2.5$) there are only local synchronized groups, and clearly only very 
close oscillators influence each other. 

\begin{figure}[H] \centering 
\includegraphics[width=.7\textwidth,angle=0]{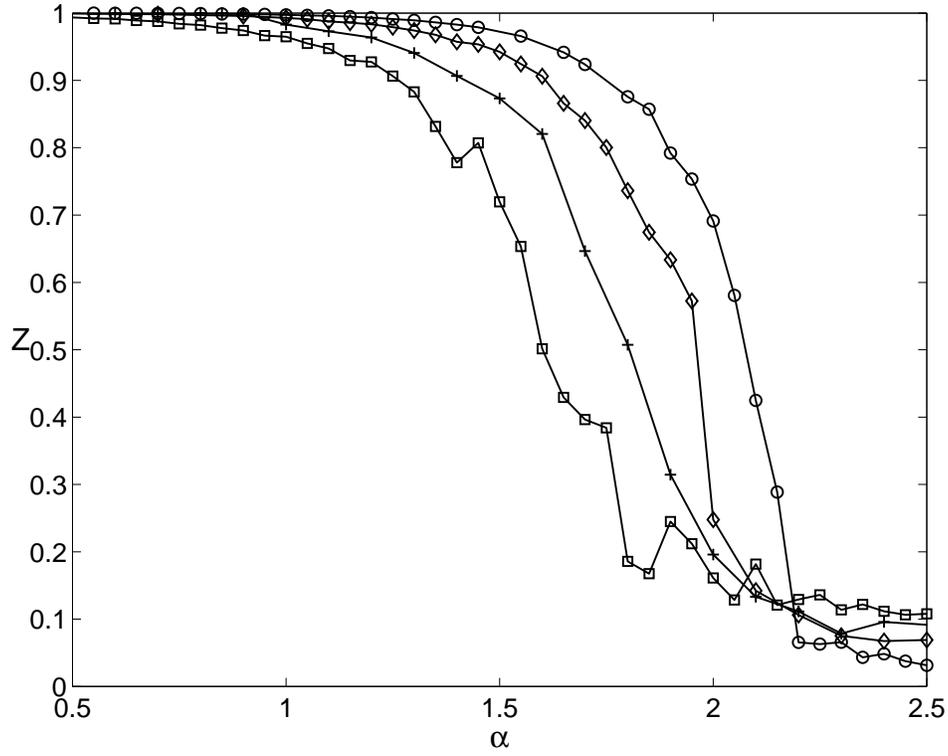}
\caption{Mean field $Z$ as a 
function of $\alpha$ for $d=2$ for $l\times l$ lattices. $l$ values are 7 
(squares), 10 (pluses), 15 (diamonds), 20 (circles).
The coupling constant is $K=0.1$, and results were averaged over more 
configurations,
especially for low $l$.} \label{fig:az2d} \end{figure}

The transition can be observed also in two dimensions (see Fig.\ 
\ref{fig:az2d}). 
Simulations were conducted on a square lattice with the coupling constant 
 set to $K=0.1$, otherwise all other parameters are the same as in $d=1$. As in the 
previous case, the transition becomes sharper as one increases $N$.

\begin{figure}[H] \centering 
\includegraphics[width=.7\textwidth,angle=0]{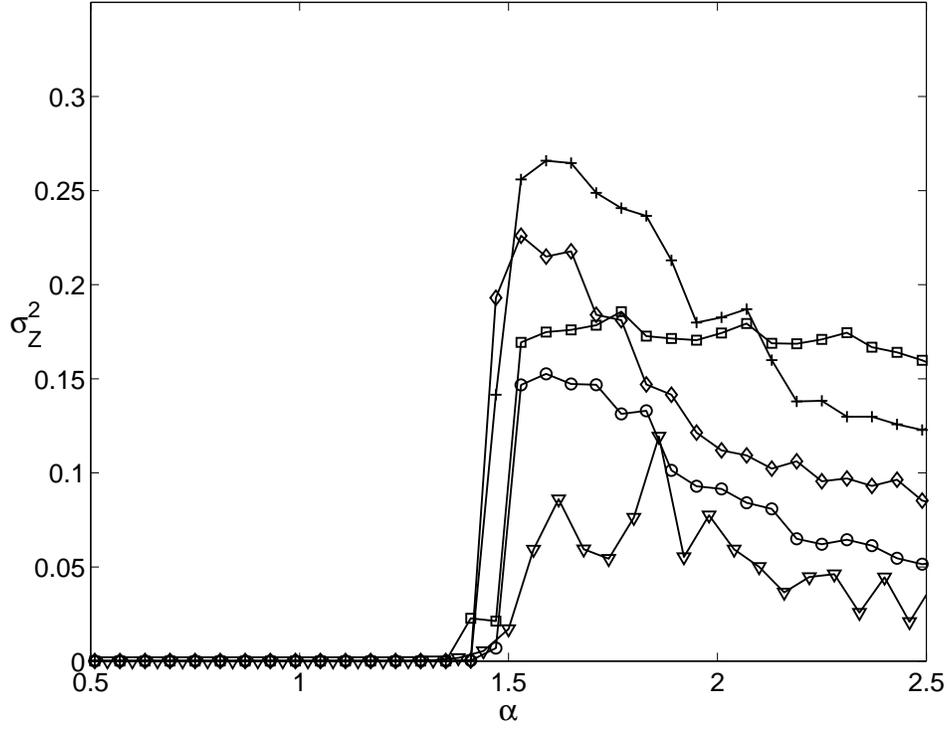} 
\caption{The standard 
deviation of the order parameter as a function of the $\alpha$ exponent for 
$K=1$. Symbols as in Fig. \ref{fig:az1d}.\label{fig:figq2}} \end{figure}

The phenomenon we present here is reminiscent of thermodynamic phase 
transitions in several features. First, the transition from unsynchronized 
state to a synchronized one breaks the original rotational symmetry of phases. 
Second, a 
phenomenon similar to the divergence of fluctuations close to the transition 
point can be observed also in this system, considering the standard deviation 
of the order parameter time-series (see Fig.\ \ref{fig:figq2}).

\begin{figure}[H] \centering              
\includegraphics[width=.7\textwidth,angle=0]{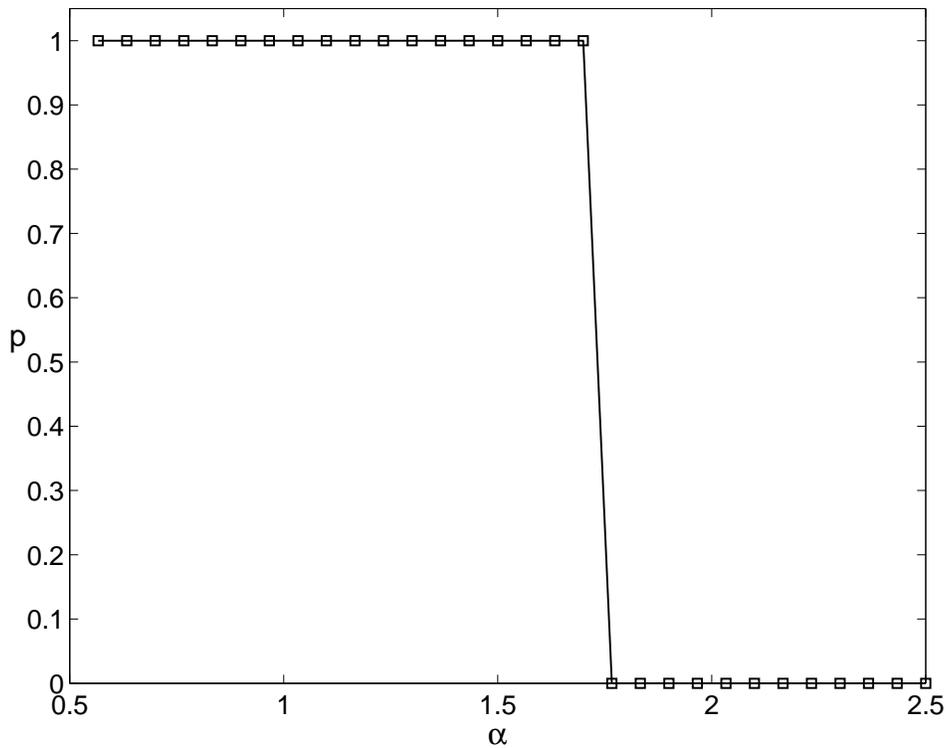} 
\caption{The fraction $p$ of oscillators locked at the mean frequency for
$K=1$, $N=200$.\label{fig:figp}} \end{figure}

It is interesting to compare the behaviour of $Z$
(Fig. \ref{fig:az1d}) with 
the plot obtained looking at the fraction of oscillators locked at the mean 
frequency (Fig.\ \ref{fig:figp}), used in the full locking approach. Although 
in the second case a discontinuity appears for finite $N$, one cannot 
distinguish between the transition region (with some unlocked oscillators but 
a strong mean field) and the region at high $\alpha$ (with developed 
incoherence and a low mean field).

\begin{figure}[H] \centering
\includegraphics[width=.7\textwidth,angle=0]{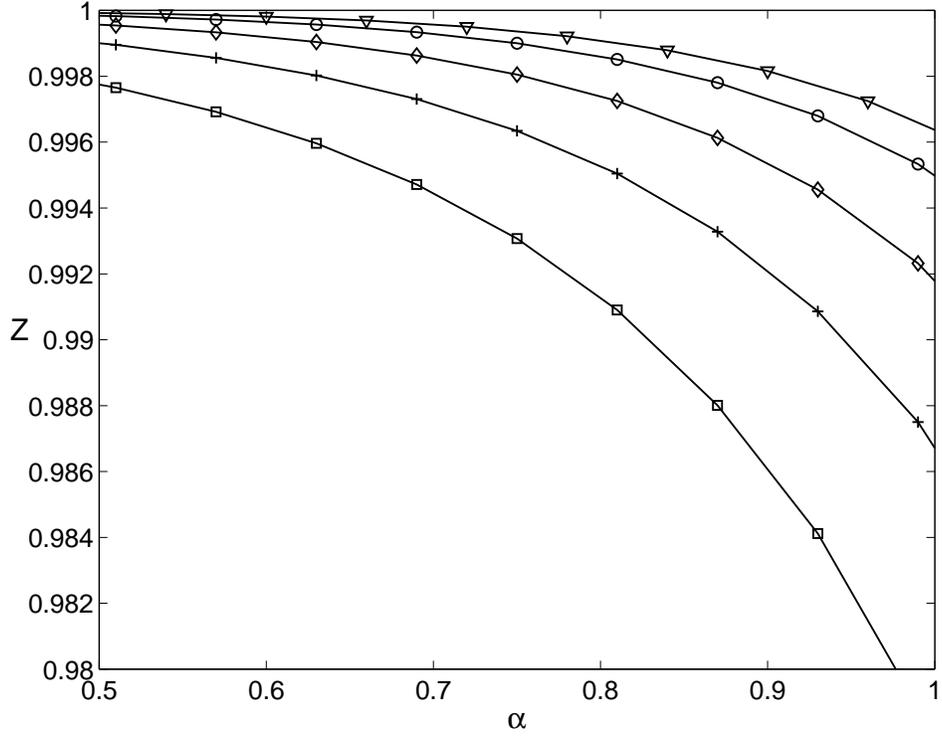} 
\caption{Enlargement of Fig. \ref{fig:az1d}. Synchronization region.}
\label{fig:figzoom1} \end{figure}

\begin{figure}[H] \centering
\includegraphics[width=.7\textwidth,angle=0]{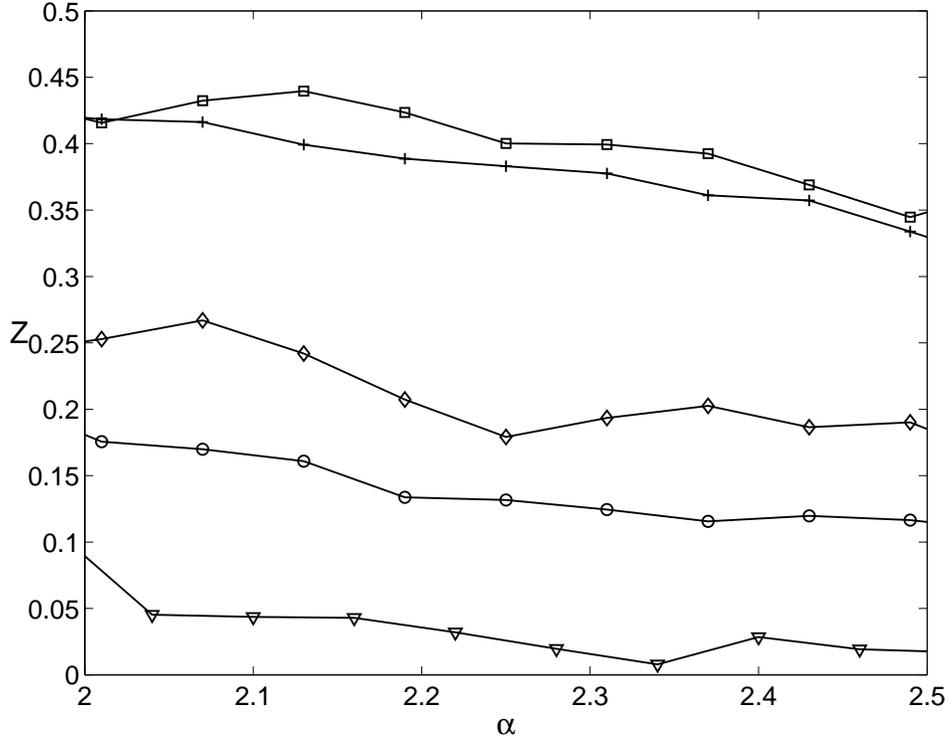} 
\caption{Enlargement of Fig. \ref{fig:az1d}. Incoherence region.}
\label{fig:figzoom2} \end{figure}

Another effect that was missed looking at full locking only is the dependence 
of the mean field on $N$. From Fig.\ \ref{fig:figzoom1} one can see that
at low 
$\alpha$ $Z$ increases with $N$, since the oscillators rotate closer to
each 
other (i.e, with smaller phase differences). We remark that this can happen 
even if the systems is not full locked. The effect on $Z$ is the opposite
for 
incoherence (high $\alpha$). In this case, the decrease in the mean field is 
due to the fact that for high $N$, statistical fluctuations are reduced when 
averaging the uncorrelated phases over a larger number of the oscillators
(Fig. \ref{fig:figzoom2}).

\begin{figure}[H] \centering 
\includegraphics[width=.7\textwidth,angle=0]{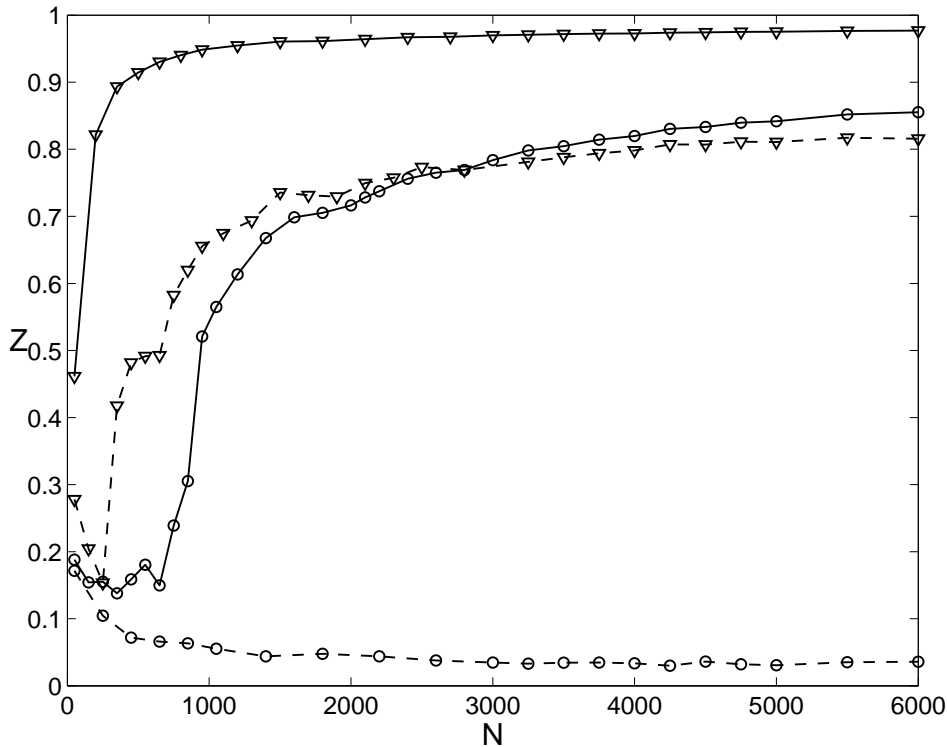}
\caption{Size dependence of order parameter for $\alpha$ below (continuous 
line, $\alpha=0.9$) and above (dashed line, $\alpha=1.1$) the critical
value $\alpha=d=1$. 
The coupling is $K=0.2$ (triangles) and 0.1 (circles).}
\label{fig:figq3} 
\end{figure}

The dependence of $Z$ over $N$ is also relevant when looking for a phase 
transition. In fact, it points to a discontinuity in the plot
$Z=Z(\alpha)$ 
for $N$ going to infinity, as Figs.\ \ref{fig:az1d} and \ref{fig:az2d} 
suggest. 
In order to detect the phase transition, we perform simulations aimed at 
studying this effect. Results are plotted in Fig.\ \ref{fig:figq3}. 
Considering 
as in Sec. \ref{sec:kur} the convergence properties of the coupling term,
and 
calling $d$ the number of dimensions of the lattice, they can be interpreted 
as follows.

For $\alpha>d$ the coupling term is bounded for $N$ going to infinity. Hence, 
for $K$ sufficiently small, the system is incoherent and the mean field 
approaches 0 when $N$ increases. For $\alpha\le 1$ the coupling term is 
unbounded for $N$ going to infinity (i.e., may diverge in some regions of the
phase space). Hence, for any (fixed) $K$, however 
small, the mean field asymptotically approaches 1 for $N$ going to infinity. In 
mathematical terms:

\begin{equation} \lim_{N\to \infty} \lim_{K\to 0^+} Z(\alpha)=\left\{ 
\begin{array}{ll} 1 & \mbox{if $\alpha \le d$} \\ 0 & \mbox{if $\alpha>d$ }, 
\end{array} \right. \label{eq:limlim2} \end{equation}
This double limit gives a compact and meaningful result on the synchronization 
properties of a system in relation to the decay of the coupling signal.

\section{Population of oscillators with thermal noise}

\label{sec:thermal}

The analogy with thermodynamic phase transitions can actually be developed 
further. In this Sec. we show that a similar transition takes place when the 
natural frequencies are equal, and randomness is introduced with thermal 
noise. The population of oscillators becomes then very close to a
Heisenberg system with given temperature.

We rewrite the natural frequencies in the form \begin{equation} 
\omega_i(t)=\omega_0+\xi_i(t), \end{equation}
where $\xi_i(t)$ is the noise term chosen from some distribution. We
choose $\xi_i(t)$ as a 
Gaussian distribution, such that $\overline{\xi_i(t)} =0$ and $\overline{ 
\xi_i(t)\xi_j(t')}=2D\delta_{ij}\delta(t-t')$. It is clear that for 
every $t$ the natural frequencies $\omega_i$ have the same distribution as in 
the above discussed model. However, the realization of this distribution 
changes at each instant of time. The equations of motion for the oscillators thus 
become:

\begin{equation} \dot\phi_i=\omega_0+K\sum_{j\ne i}\frac{1}{r_{ij}^{\alpha}} 
\sin(\phi_j-\phi_i) +\xi_i(t). \end{equation}

\begin{figure}[H] \centering
\includegraphics[width=.7\textwidth,angle=0]{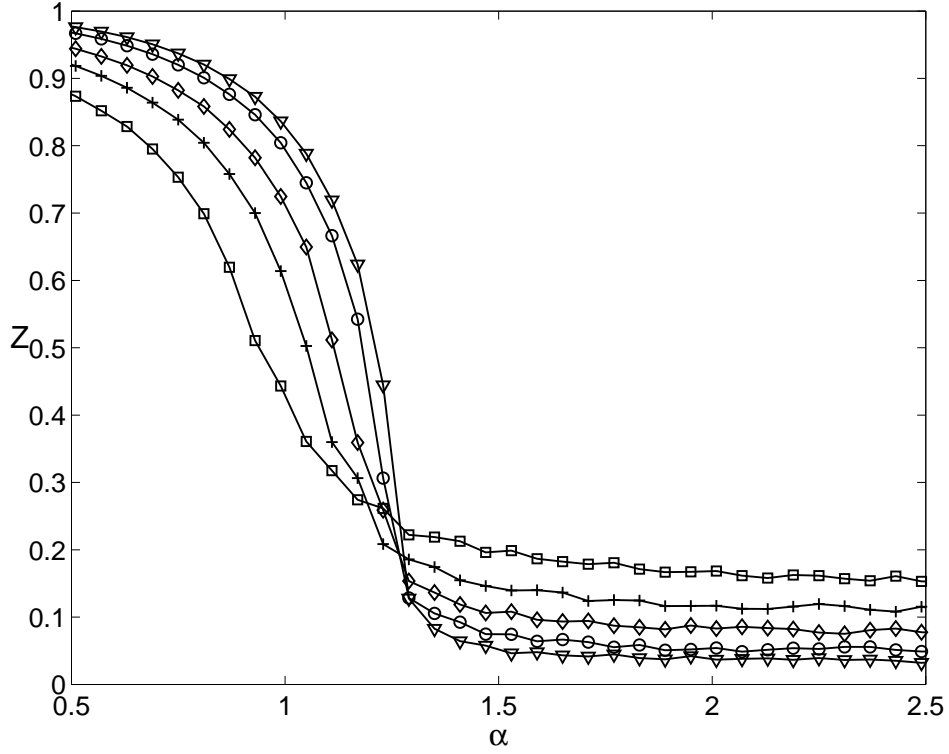} 
\caption{Average of the order parameter as a function of $\alpha$ for 
identical oscillators with noise in $d=1$. Simulation results for $N$=50, 100, 
200, 500, and 1000 (symbols are squares, pluses, diamonds, circles,
and down triangles respectively). The coupling is $K=1.0$.} 
\label{fig:noise} \end{figure}

\begin{figure}[H] \centering 
\includegraphics[width=.7\textwidth,angle=0]{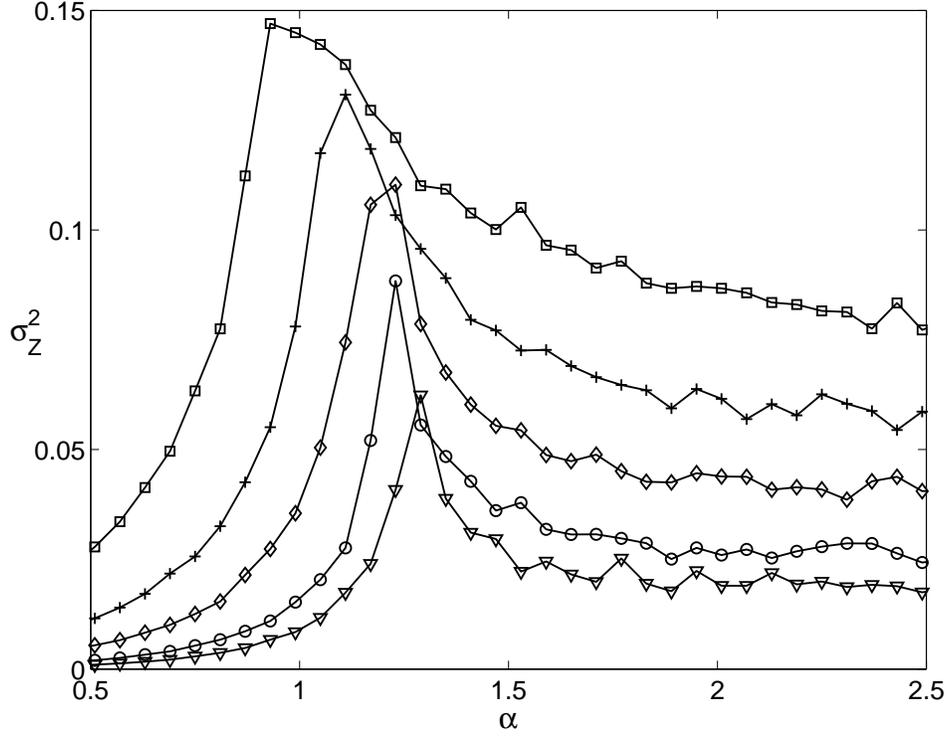}
\caption{Standard deviation 
of the order parameter as a function of $\alpha$ for identical oscillators 
with noise in $d=1$. Symbols are the same as in Fig. \ref{fig:noise}. 
} \label{fig:noisestd} 
\end{figure}

\begin{figure}[H] \centering 
\includegraphics[width=.7\textwidth,angle=0]{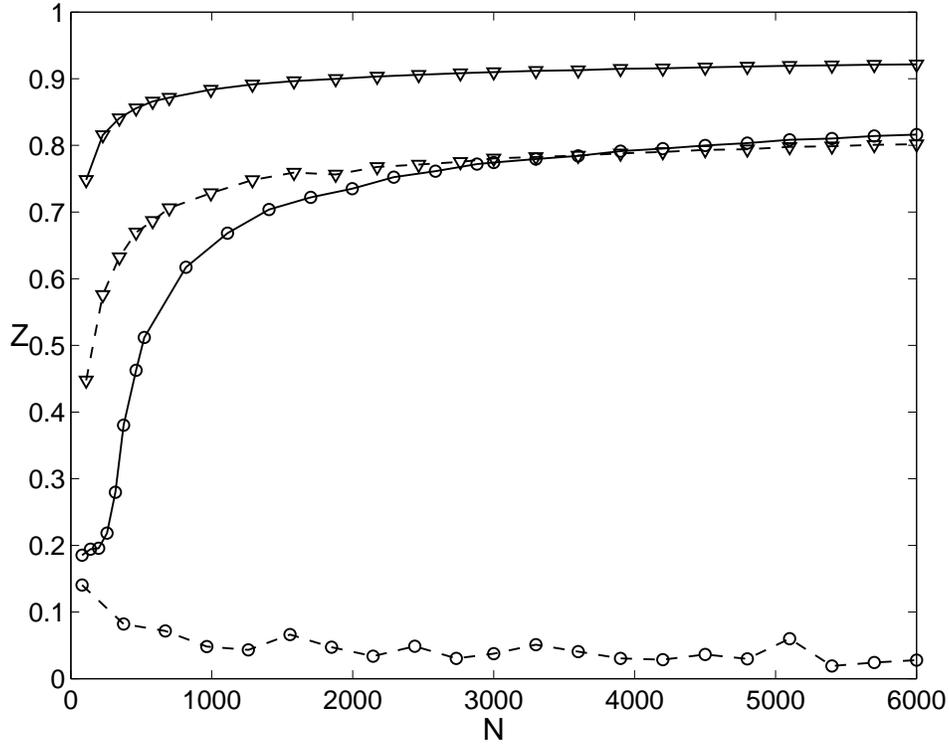}
\caption{Size dependence of order parameter for identical oscillators
with noise for $\alpha$ below (continuous 
line) and above (dashed line) the critical value $\alpha=d=1$. 
The coupling is $K=1.0$ (triangles) and $0.5$ (circles).} \label{fig:figt3} 
\end{figure}

Our simulations show that the phase transition from synchronized to
unsynchronized state described in the previous Sec. takes place
in this arrangement also
(Figs. \ref{fig:noise} and \ref{fig:noisestd}). Besides the transition,
the size effects appear to remain valid in this case. For high $K$, 
synchronization does not depend on the relation between  $\alpha$ and the 
lattice dimensions (Fig. \ref{fig:figt3}, triangles). However, for low 
coupling there is indication that the system synchronizes only if 
$\alpha\leq d$ (Fig. \ref{fig:figt3}, circles).

\section{Discussion} \label{sec:disc}

We have considered synchronization of phase oscillators on a lattice, looking 
at critical levels of spatial decay in the interaction. Especially, since the 
coupling term considered is not normalized, we studied the effect  of
changing the population size on 
synchronization. We examined the system by investigating full locking 
and the mean field. The two approaches have appeared to be 
complementary. The criterion of full locking allowed to precisely define the 
boundary of complete synchronization for finite population sizes. 
It gives however a very strong condition, requiring all oscillators to be 
exactly locked. For small coupling, this condition is not useful. If the 
coupling constant is small, synchronization can still be enhanced, below a 
critical $\alpha$ value, enlarging the size of the population: but some of the 
oscillators remain unlocked. We thus studied the system from a different 
point of view, that is, looking at the behaviour of the mean field. In that 
case, a transition point is not strictly defined for finite population size. 
However, in the limit of infinite number of oscillators one can look, in 
analogy with thermodynamics, to the mean field as an order parameter and thus 
find critical values of the parameters where discontinuities appear. As one 
may expect, the value of the decay exponent equal to the number of lattice 
dimensions is then a good candidate for a transition point. At that value and 
below in fact, the coupling term is unbounded for an infinite size. That was 
supported by numerical simulations. 
As we pointed out, this gives a robust result for real systems: knowing only 
the number of lattice's dimensions and the decay in space of the coupling 
signal, one can predict if enlarging the size of the system eventually results 
in synchronization or not, even for arbitarily weak coupling constant.

We finally considered a system of oscillators in which the diversity is given 
not by fixed natural frequencies, but by noise. The notion of full locking is 
not useful for this system, but the mean field approach can be carried out, 
and suggests the same features and critical point, at a decay exponent equal 
to the lattice's number of dimensions. Beside showing the robustness of the 
result, this last result is promising for an analytical treatment of the 
problem. In fact, the system with thermal noise is close to a Heisenberg
spin-system. 
We remark that an analytical investigation would be important. In fact, 
simulations are very time consuming, and allowed us to have indications of the 
phenomenon in a relatively small parameter region. Especially, the analysis of 
higher dimensional lattices, as well as lower coupling strength, would be 
interesting.

\begin{acknowledgments} F. d'Ovidio thanks E. Mosekilde and S. De Monte for the 
fruitful discussions. 
M. Mar\'{o}di and T. Vicsek 
are grateful to A. Czir\'{o}k and Z. N\'{e}da for many helpful discussions.
Support from OTKA (Hungary, Grant No: T034995) is
acknowledged. Some of the simulations were 
performed on the parallel cluster of MTA SZTAKI. \end{acknowledgments}

\bibliography{artbib}

\end{document}